# Visualizing the mixed bonding properties of liquid boron with high resolution Compton scattering

J.T. Okada[1,2], P.H.-L. Sit[3], Y. Watanabe[4], B. Barbiellini[5], T. Ishikawa[1], Y.J. Wang[5], M. Itou[6], Y. Sakurai[6], A. Bansil[5], R. Ishikawa[4], M. Hamaishi[4], P.-F. Paradis[1], K. Kimura[7], T. Ishikawa[8], and S. Nanao[1,]

[1]*Institute of Space and Astronautical Science, Japan Aerospace Exploration Agency, Tsukuba, Ibaraki 305-8505, Japan*
[2]*PRESTO, JST, 4-1-8 Honcho, Kawaguchi, Saitama 332-0012, Japan*
[3]*School of Energy and Environment, City University of Hong Kong, Hong Kong*
[4]*Institute of Industrial Sciences, The University of Tokyo, Meguro-ku, Tokyo, 153-8505, Japan*
[5]*Department of Physics, Northeastern University, Boston, Massachusetts 02115, USA*
[6]*Japan Synchrotron Radiation Research Institute, SPring-8 Sayo-cho, Hyogo 679-5198, Japan*
[7]*Department of Advanced Materials Science, The University of Tokyo, Kashiwa, Chiba 277-8561, Japan*
[8]*RIKEN SPring-8 Center, Sayo-cho, Sayo-gun, Hyogo 679-5148, Japan*

Bonding characteristics of liquid boron at 2500K are studied by using high resolution Compton scattering. An excellent agreement is found between the measurements and the corresponding Car-Parinello molecular dynamics simulations. Covalent bond pairs are clearly shown to dominate in liquid boron along with the coexistence of diffuse pairs. Our study reveals the complex bonding pattern of liquid boron, and gives insight into the unusual properties of this high temperature liquid.

Boron (B), which is abundant on earth, is a technologically important material because of its large degree of hardness, its low density and high melting temperature [1]. B-rich compounds display remarkable properties: for example, $MgB_2$ shows high temperature superconductivity at 39 K [2], cubic-BN is the second hardest material after diamond, and $B_4C$ is the third hardest material [3]. Recent high-pressure experiments provide evidence of exciting new characteristics of elemental B: superconductivity at 11.2 K (a relatively high transition temperature for a simple element) under a pressure of 250 GPa [4], and the existence of a new B phase γ-$B_{28}$ between 19 and 89 GPa [5]. These recent findings have renewed interest in fundamental properties of elemental B. Interestingly, B belongs to the group of elements situated between metals and insulators in the periodic table. B possesses three outer electrons like aluminum and as a result it can support a metallic state, but the B atom is smaller, and therefore its atomic potential holds the electrons more tightly like an insulator. This balance between metallic and insulating tendencies could be altered via pressure, temperature, or impurities. For example, elements in the column on the right of B in the periodic table, such as carbon [6], silicon (Si), and germanium, which are semiconductors in the solid state, transform into metals on melting. Some theoretical studies suggest that B could also acquire some metallic character on melting [7,8,9]. Although transport experiments on liquid (*l*-) B have indicated the survival of a semiconducting behavior [10], the question of whether *l*-B is a metal or not remains controversial and unsolved.

Until very recently, it was difficult to carry out experiments on liquid B since it is extremely reactive with all known container materials. The recent advent of levitation techniques has made it possible to handle samples without contamination and to carry out measurements of physical properties of *l*-B in the temperature region down to the supercooled state. Unusual properties of *l*-B have thus been revealed. For instance, B contracts by nearly 3% on melting [11] like Si and Germanium [12] while most materials expand on melting. Dynamical properties of l-B differ from that of simple monoatomic liquids [13]. Moreover, viscosity of *l*-B increases substantially with decreasing temperature in the supercooled state, suggesting the possible freezing of an amorphous percolating covalent network [14]. In this Letter, we reveal the complex bonding characteristics of *l*-B through Compton scattering experiments in levitation combined with first principles calculations, gaining insight into the unusual properties of *l*-B.

X-ray Compton scattering [15] is an ideal technique for probing the ground state wave functions in materials [16-19]. The Compton profile $J(p_z)$ obtained from the measurement is given within the impulse approximation by [20],

$$J(p_z) = \iint n(\mathbf{p}) dp_x dp_y , \quad (1)$$

where $\mathbf{p} = (p_x, p_y, p_z)$ and $n(\mathbf{p})$ is the electron momentum density given by [15]

$$n(\mathbf{p}) = \sum_j \left| \psi_j(\mathbf{r}) \exp(-i\mathbf{p}\cdot\mathbf{r}) d\mathbf{r} \right|^2 , \quad (2)$$

where $n_j$ is the electron occupation of $j^{th}$-state $\Psi_j(\mathbf{r})$. The area under the Compton profile gives the total number of electrons N:

$$\int_{-\infty}^{\infty} J(p_z) dp_z = N . \quad (3)$$

The Compton technique can also provide a novel spectroscopic window on the liquid state. Since no charged particles entering or leaving the sample are detected, the technique is a genuinely bulk probe, which is not complicated by surface effects present in photoemission or electron scattering experiments.

Compton profiles of polycrystalline β-rhombohedral (β-) B (300 K) and molten B (2500

K) were measured by high-energy (116 keV) inelastic x-ray scattering at the BL08W beamline of SPring-8 [21]. The energy spectrum of Compton scattered X-rays was converted to the Compton profile with a momentum resolution of 0.16 a.u. The data processing to deduce the Compton profile from the raw energy spectrum consists of following steps: background subtraction, energy dependent corrections for the Compton scattering cross section, the absorption of incident and scattered x-rays in the sample, consideration of efficiency of the analyzer and the detector, and corrections for double scattering events: see [22]. Background in the measured profile is suppressed in the levitation technique used since no material is present in the immediate vicinity of the sample at the x-ray scattering center.

Molten B is highly reactive with most crucibles. In order to hold the sample without contamination, a high temperature electrostatic levitator (HTESL) was used [23]. The detail of the HTESL for SPring-8 experiments is described in [24]. A polycrystalline sample (β-B, 99.9% purity) was used to obtain a spherical averaging Compton profile. The sample was heated and melted using the focused radiation of three 100 W semiconductor laser beams emitting at 808 nm. Temperature was controlled with an accuracy of 15 K and measured by pyrometry at wavelengths of 0.90 and 0.96 nm.

Density functional theory (DFT) [25][26] and Car-Parrinello molecular dynamics (CPMD) simulations [27] were performed with the Quantum ESPRESSO package [28] within the framework of the density-functional theory using the generalized gradient approximation. We employed ultrasoft pseudopotentials with plane-wave expansion of the Kohn-Sham wave functions and charge density up to kinetic energy cutoff of 25 Ry and 200 Ry, respectively. For the $l$-B simulation, spin-unpolarized simulations were done with 256 B atoms in a simple cubic supercell. A time step of 0.097 fs was used with the fictitious mass of electron set to 300 a.u., which is within the suggested range for accurate CPMD simulations. A structural optimization calculation was performed on solid B for the α- and β-rhombohedral phases at 0 K with 144 and 210 atoms, respectively. The simulations were carried out at the experimental densities of the α- and β- phases of 2.46 g/cm$^3$ and 2.35 g/cm$^3$, respectively [1]. The *l*-B simulation was performed at 2500 K at the experimental density of 2.14 g/cm$^3$ [10]. The Nose-Hoover thermostat on ions was applied in the liquid simulations [29][30]. We also used an additional Nose-Hoover thermostat on electrons in the liquid simulation due to the metallic character of the system [31]. The liquid simulation lasted for around 8.5 ps. Compton profile of *l*-B was calculated after thermalization from the average of eight uncorrelated configurations, equally spaced by 1 ps. Both profiles were spherically averaged and convoluted with a Gaussian to match the 0.16 a.u. experimental resolution. The valence-electron Compton profiles of β-B (300 K) and *l*-B (2500 K), obtained by subtracting the theoretical core-electron profile $J_c(p_z)$ [32] from the experimental profiles, are presented in Fig. 1. The theoretical core-electron profiles are based on the free-atom Hartree-Fock simulations in which the 1$s^2$ shell is treated as core electrons.

Figure 2 presents the difference between the Compton profiles for the solid and liquid phases [$\Delta J(p_z) = J_{\beta\text{-B}}(p_z) - J_{l\text{-B}}(p_z)$]. The CPMD calculated results are obtained by taking the difference of the liquid Compton profile and the spherically averaged solid Compton profile. These theoretical predictions are seen to agree extremely well with the experimental results demonstrating the high reliability of the bonding properties of *l*-B obtained through the CPMD simulations. The simulations also display an excellent agreement with the corresponding pair correlation function measured by x-rays [8].

We analyzed the bonding characteristics of *l*-B through CPMD simulations in terms of maximally-localized Wannier functions (MLWFs) [33, 34] where the MLWFs were obtained via a unitary transformation of Kohn-Sham wave functions. The spread of MLWFs is an indicator of the extent to which the associated electronic states are localized. An MLWF with a small spread identifies a localized covalent bond pair. In contrast, an MLWF with a large spread indicates the presence of a delocalized electronic state (we refer to such electron pairs as diffuse pairs), which will contribute to metallic properties.

Figure 3(a) shows the Wannier function spread distributions (WFSDs) in B. All the electron pairs in the $\alpha$-B phase and the majority of electron pairs in $\beta$-B phase are seen to be covalent, and our MLWFs are consistent with the results obtained by Ogitsu *et al* [35]. The spread of about 3 Å$^2$ in the WFSD of $\beta$-B reflects the presence of disorder in the form of vacancies and interstitials, which has been incorporated in our CPMD simulations [35]. Although the WFSD of *l*-B displays a long tail extending to 11.2 Å$^2$, it essentially overlaps with that of $\beta$-B, indicating a substantial contribution in *l*-B from covalent bond pairs. The long tail in the WFSD of *l*-B, on the other hand, shows the presence of diffuse (metallic) bond pairs. Using the criteria described in Ref. [36] that the boron-boron bond length is 2.35 Å (compared to 3.1 Å in Si), we estimate that on the average there are 66%, 33%, and 1% covalent, diffuse and lone pairs, respectively.

It is clear thus that two distinct bonding species, covalent and metallic, both exist together in *l*-B. Such a coexistence of covalent and diffuse bond pairs is also observed in *l*-Si [24]. For comparison, the computed WFSD of Si is shown in Fig. 3(b). Although the WFSD of *l*-Si shows a long tail like *l*-B, the overlap between solid (*s*)-Si and *l*-Si is seen to be rather small, indicating a reduced covalent character: the percentage of covalent, diffuse, and lone pair bonds in *l*-Si is estimated to be 17%, 83%, and less than 1%, respectively [24].

We turn now to address the question whether or not $l$-B is a metal. Note in Fig. 3(a) that the center of the WFSD in $l$-B is fairly close to the WFSD of $s$-B. In contrast, the center of the WFSD for $l$-Si in Fig. 3(b) is well separated from the WFSD of $s$-Si. Table 1 compares the average and standard deviation of the WFSD of B and Si. The average value can be used to monitor the covalent character of the electronic system: the smaller is the average, the larger the covalent character. As expected, the average of the WFSD is much smaller in $l$-B compared to $l$-Si, consistent with the fact that the melting temperature of B is higher than that of Si, and B is harder than Si. Moreover, the population of covalent pairs in $l$-B is much larger than in $l$-Si. Although some diffuse pairs with large Wannier spreads are found in $l$-B, the overall contribution of diffuse pairs is quite small. The covalent character is thus much more pronounced in $l$-B than in $l$-Si, allowing us to conclude that $l$-B possesses a mixed character, which is dominated by semiconducting properties coexisting with weak metallic characteristics. This could coincide with the transport experiment on $l$-B [10].

Figure 4 shows a snapshot from the MD simulation carried out on $l$-B and demonstrates that bonding is mostly covalent. A network of covalent bonds is seen to percolate throughout the liquid and icosahedral clusters could not be existed. These covalent bonds possess a highly dynamical nature and change rapidly on a time scale of a few tens of femtosecond, consistent with previous studies [13].

To summarize, we have carried out high resolution Compton scattering measurements on β- and $l$-B and interpreted the results in terms of parallel CPMD simulations. Experimental results are found to agree exceptionally well with the corresponding theoretical predictions, allowing us to estimate that *l*-B contains 66%, 33% and 1%, covalent, diffuse and lone pair bonds, respectively. The covalent bond pairs thus give the largest contribution. We conclude that *l*-B possesses a mixed character dominated by semiconducting properties admixed with some metallic states. Although many spectroscopic techniques are well developed for analyzing the electronic structures of crystals, this is not the case when it comes to high temperature liquids. The experimental and theoretical approaches discussed in the present study provide new tools of wide applicability for investigating the nature of bonding in high temperature liquids. For instance, WFSDs could also provide useful information for liquid tellurium despite the fact that the treatment of higher atomic systems is more difficult [37].

We acknowledge important discussions with S. Kaprzyk. Compton profile measurements were performed with the approval of JASRI (Proposals No. 2007B1235). The work at JAXA was supported by JST, PRESTO and Grants-in-Aid for Scientific Research KAKENHI from MEXT of Japan under Contracts No. 16206062 and 26709057. The work was supported by the Start-up Grant No. 7200397 at the City University of Hong Kong, the US Department of Energy, Office of Science, Basic Energy Sciences Grants No. DEFG02-06ER46344 at Princeton University and No. DEFG02-07ER46352 and No. DE-SC0007091 (CMCSN) at Northeastern University, and benefited from the allocation of time at NERSC and NU's Advanced Scientific Computation Center.

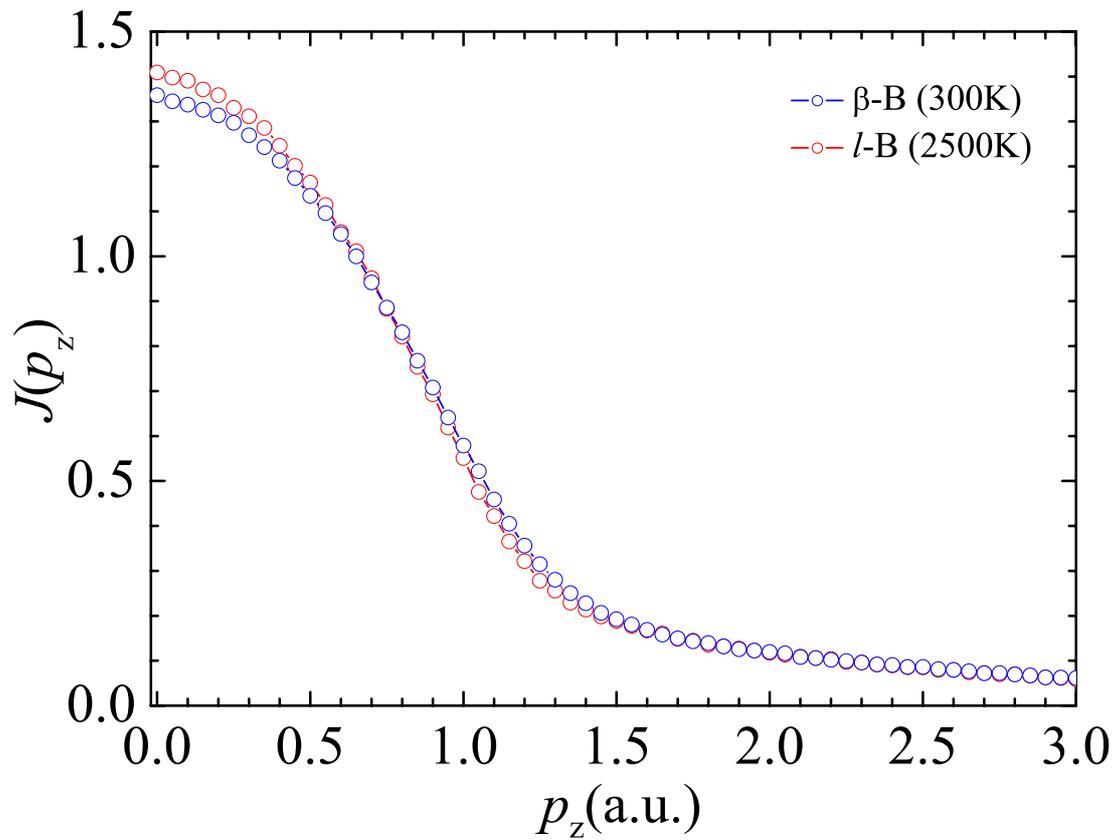

Figure 1. (color online) Experimental valence Compton profile of solid β-rhombohedral (blue open circles) and liquid B (red solid circles). The error bars are less than the size of symbols.

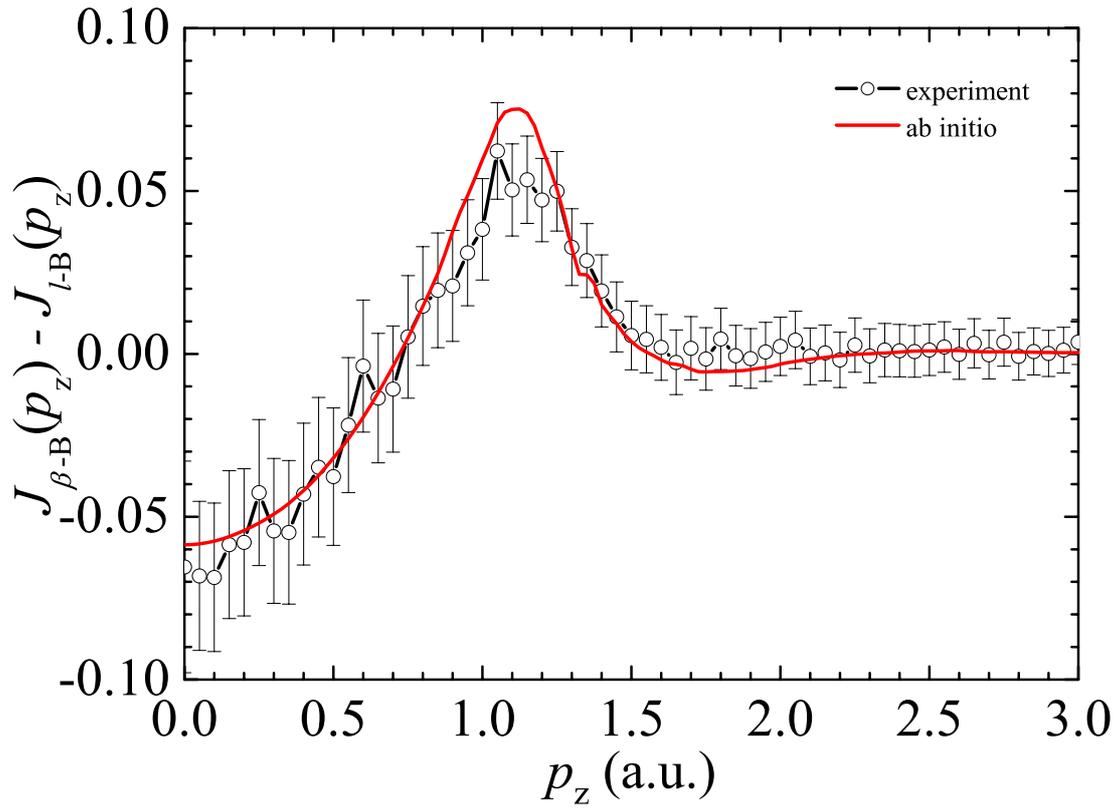

Figure 2. (color online) Difference between the Compton profiles of solid (β-B) and liquid phases of B [$\Delta J(p_z) = J_{\beta\text{-}B}(p_z) - J_{l\text{-}B}(p_z)$]: experiment (black) and CPMD simulation (red).

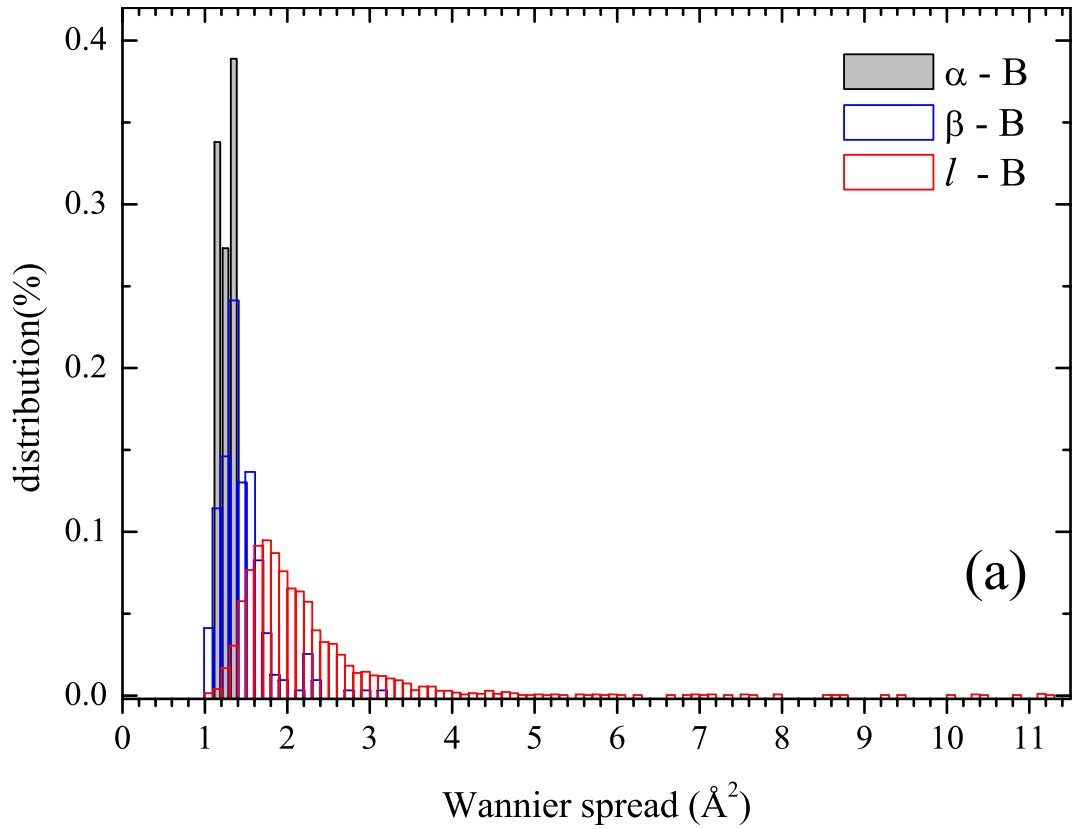

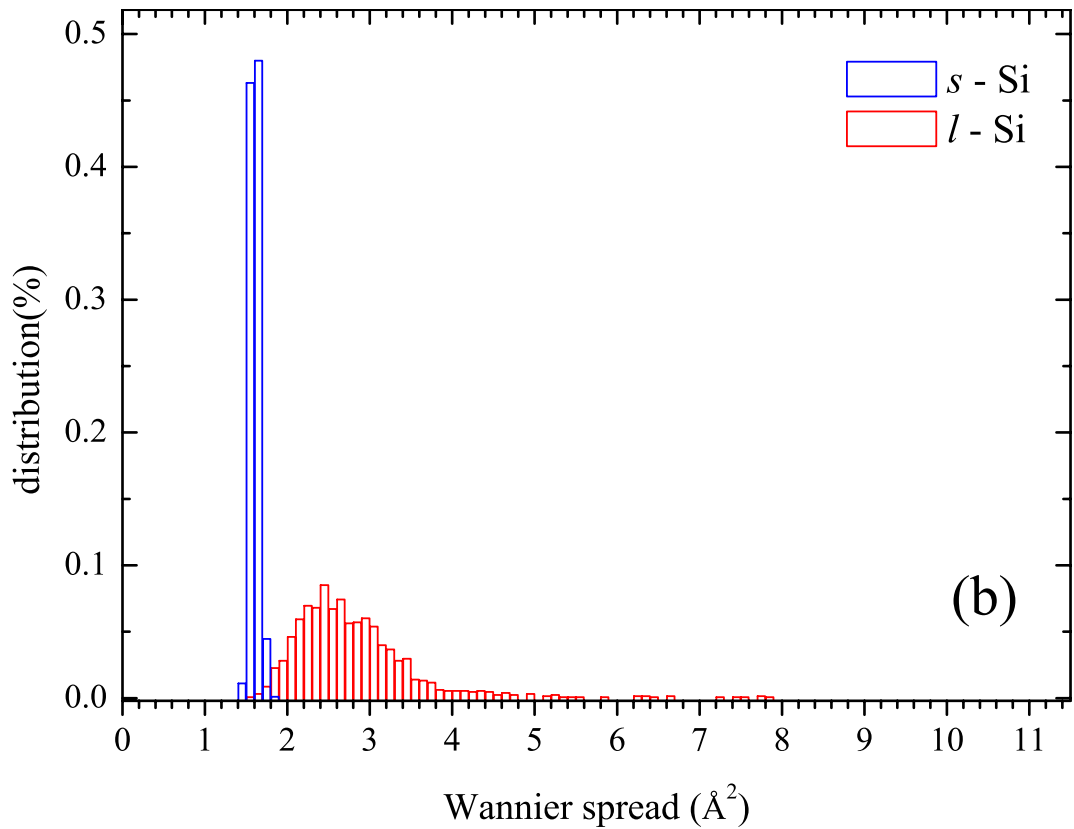

Figure 3. (color online) (a) Wannier spread distributions for various phases of B. The liquid phase distribution displays a significant tail indicating electron delocalization in *l*-B; (b) same as (a) for *l*-Si.

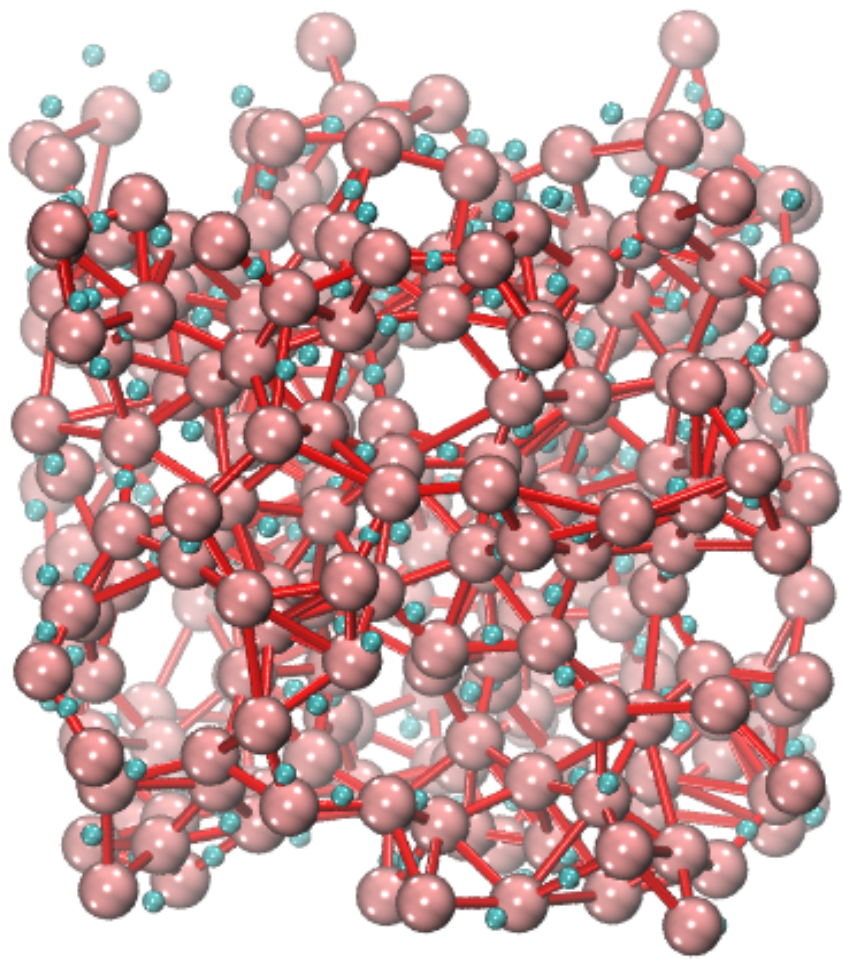

Figure 4. (color online) Snapshot of the covalent network in *l*-B at 2500 K exhibiting B atoms (pink), and covalent bond pairs (cyan). The red bonds connecting B atoms are only guides for the eye.

|  | Average (Å$^2$) | Standard deviation (Å$^2$) |
|---|---|---|
| Alpha-B | 1.243 | 0.091 |
| Beta-B | 1.438 | 0.290 |
| Liquid-B | 2.173 | 0.950 |
| Solid-Si | 1.607 | 0.052 |
| Liquid-Si | 2.796 | 0.751 |

Table I: Average and standard deviation of the Wannier spreads for different systems.